\documentstyle[11pt,gh2001-asp,twoside]{article}
\def\edcomment#1{\iffalse\marginpar{\raggedright\sl#1\/}\else\relax\fi}
\reversemarginpar

\begin{document}
\title{WIYN Integral-Field Kinematics of Disk Galaxies}


\author{Matthew Bershady$^1$, Marc Verheijen$^1$ and David Andersen$^2$}
\affil{$^1$Department of Astronomy, U. of Wisconsin--Madison, USA\\
$^2$Max Planck Institute for Astronomy, Heidelberg, Germany}

\begin{abstract}

We have completed a new fiber array, SparsePak, optimized for
low-surface-brightness studies of extended sources on the WIYN
telescope. We are now using this array as a measuring engine of
velocity and velocity-dispersion fields of stars and ionized gas in
disk galaxies from high to low surface-brightness. Here we present
commissioning data on the velocity ellipsoids, surface densities and
mass-to-light ratios in two blue, high surface-brightness, yet small
disks. If our preliminary results survive further observation and more
sophisticated analysis, then NGC 3949 \index{object, NGC 3949} has
$\sigma_z/\sigma_R\gg 1$, implying strong vertical heating, while NGC
3982's \index{object, NGC 3982} disk is substantially
sub-maximal. These galaxies are strikingly unlike the Milky Way, and
yet would be seen more easily at high redshift.


\vskip -0.25in
\end{abstract}

\section{SparsePak}

\vskip -0.05in

SparsePak contains a 76$\times$77 arcsec sparsely-packed hexagonal
grid of 75, 5-arcsec diameter fibers, which pipe light from the
``imaging'' Nasmyth focus to the versatile Bench Spectrograph on the
WIYN 3.5m telescope (Bershady et al. 2002). Seven ``sky'' fibers are
placed at $\sim$ 30 arcsec from two grid sides. The large anamorphic
demagnification in the echelle-mode of the Bench Spectrograph yields
instrumental spectral resolutions of 8000-10,000 and as high as 21,000
with SparsePak -- comparable to resolutions obtained with smaller
fibers. At the same time, SparsePak has $>90$\% throughput for
$\lambda>500$nm, and provides nearly 3$\times$ the light gathering
power over 5$\times$ the area as DensePak, the pre-existing WIYN fiber
array. These attributes make SparsePak well suited for
medium-resolution spectroscopy on extended objects down to low
surface-brightness.

\section{Disk Stellar Kinematics and Mass to Light Ratios}

\vskip -0.15in

We have mapped the velocity and velocity dispersion fields in two
blue, very high surface-brightness, yet small disks in the Ursa Major
cluster (B$-$R=0.8-1 mag, $\mu_0^{\rm obs}$(B)=19.3-19.5 mag
arcsec$^{-2}$, h$_{\rm R}$=0.9-1.7 kpc, Verheijen 1996). NGC 3982 is
nearly face-on (i=26$^\circ$), while NGC 3949 is moderately inclined
(i=54$^\circ$). As such, these photometrically similar galaxies offer
projections which enhance measurements of different components of
their velocity ellipsoids.


Each galaxy was observed in three spectral regions near 5130\AA \
(MgI), 6680\AA \ (H$\alpha$) and 8660\AA \ (CaII-triplet) with FWHM
spectral resolutions of 24, 16 and 37 km/s respectively. Stellar
absorption-line observations were taken with a single SparsePak
pointing while H$\alpha$ was contiguously mapped in three pointings at
very high S/N. The latter provided high-precision kinematic
inclinations (see Andersen \& Bershady, these proceedings). 

In the inner regions of the galaxy, the signal-to-noise in the
absorption-line spectra of individual fibers is high enough to
determine the line centroids and widths and to construct stellar
velocity fields (Figure 1). The layout of the fibers allows for
azimuthal averaging of fibers to improve signal-to-noise in the outer
regions (Figure 2). The near face-on orientation of NGC 3982, for
example, permits 18 fibers to be combined at a radius of 3.5 h$_{\rm
R}$. Before averaging, the projected rotational velocities were taken
out, using the centroids of the high signal-to-noise H$\alpha$
lines. In both galaxies this may lead to an overestimate of the inner
velocity dispersions.  Comparison of the H$\alpha$ and stellar
velocity fields in Figure 1 clearly shows the effects of asymmetric
drift inside 1.5 disk scale lengths (h$_{\rm R}$). Outside this radius
the asymmetric drift approaches zero.

\begin{figure}
\plotfiddle{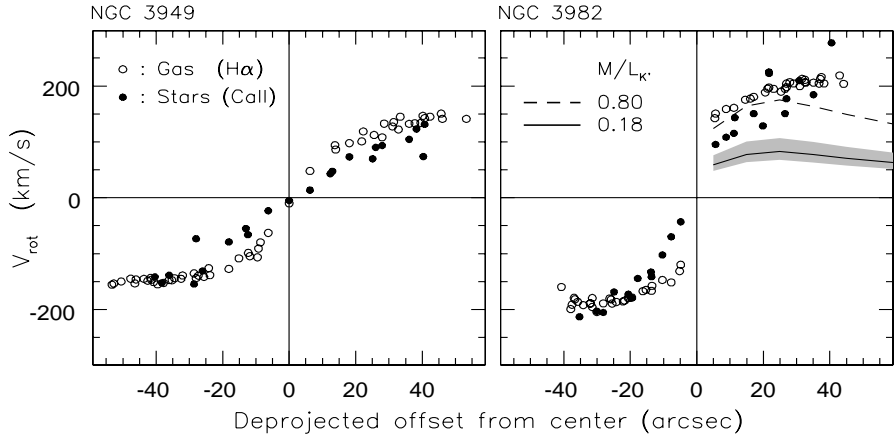}{2.0in}{0}{65}{60}{-210}{-210}
\vskip 0.05in
\caption{\hsize 5.2in \baselineskip 0.165in De-projected
position-velocity diagrams for stars (CaII-triplet) and ionized gas
(H$\alpha$) in N3949 and N3982. Note asymmetric drift, clearly visible
in the inner, steeply rising regions, approaches zero at large
radii. For N3982 we show the stellar contribution for two mass
decomposition models which include stars, gas, and a dark halo: a
maximum-disk (dashed line), and the substantially (25\%) sub-maximum
disk implied by the stellar line-widths in Figure 2 (see also
Verheijen et al. 2002).}
\vskip -0.15in
\end{figure}

\begin{figure}
\plotfiddle{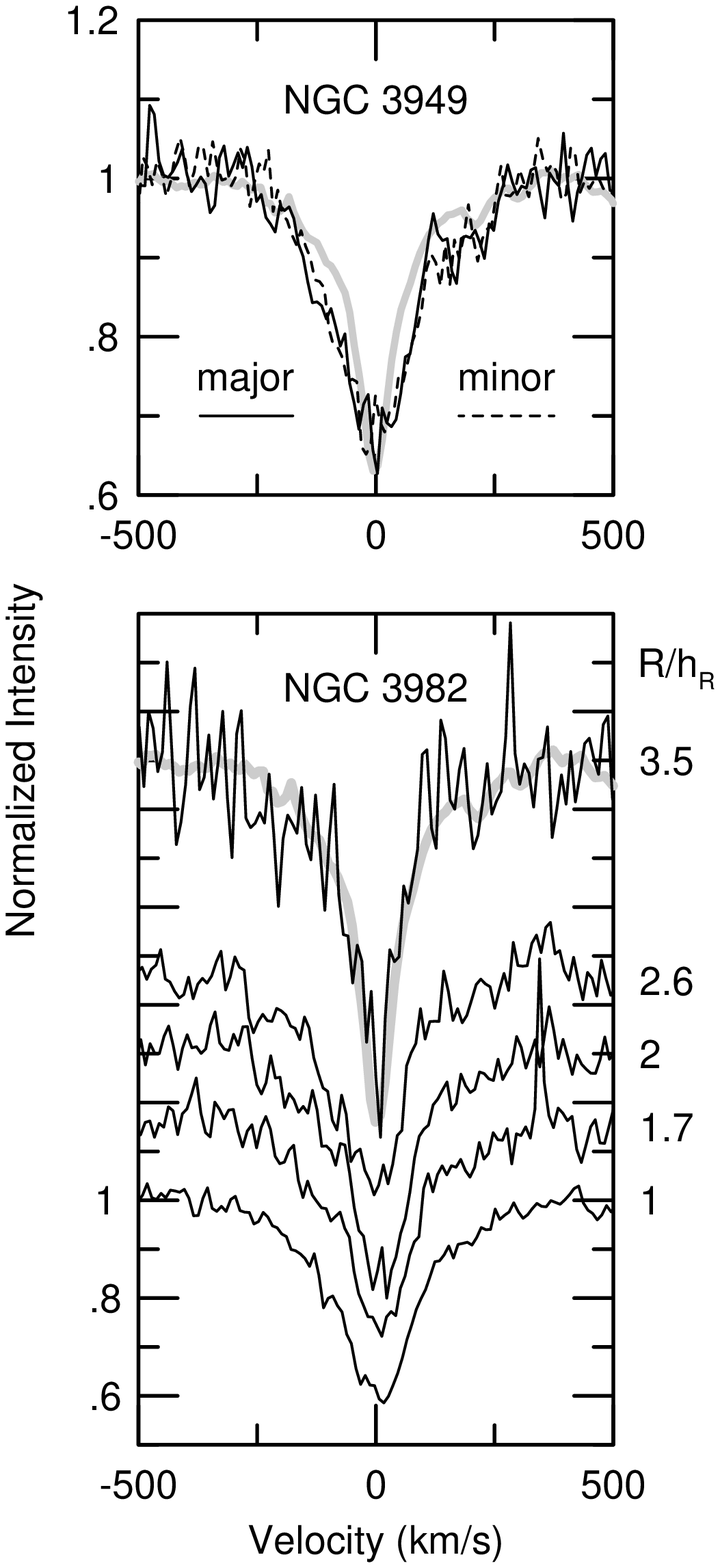}{3in}{0}{55}{55}{-200}{-155}
\plotfiddle{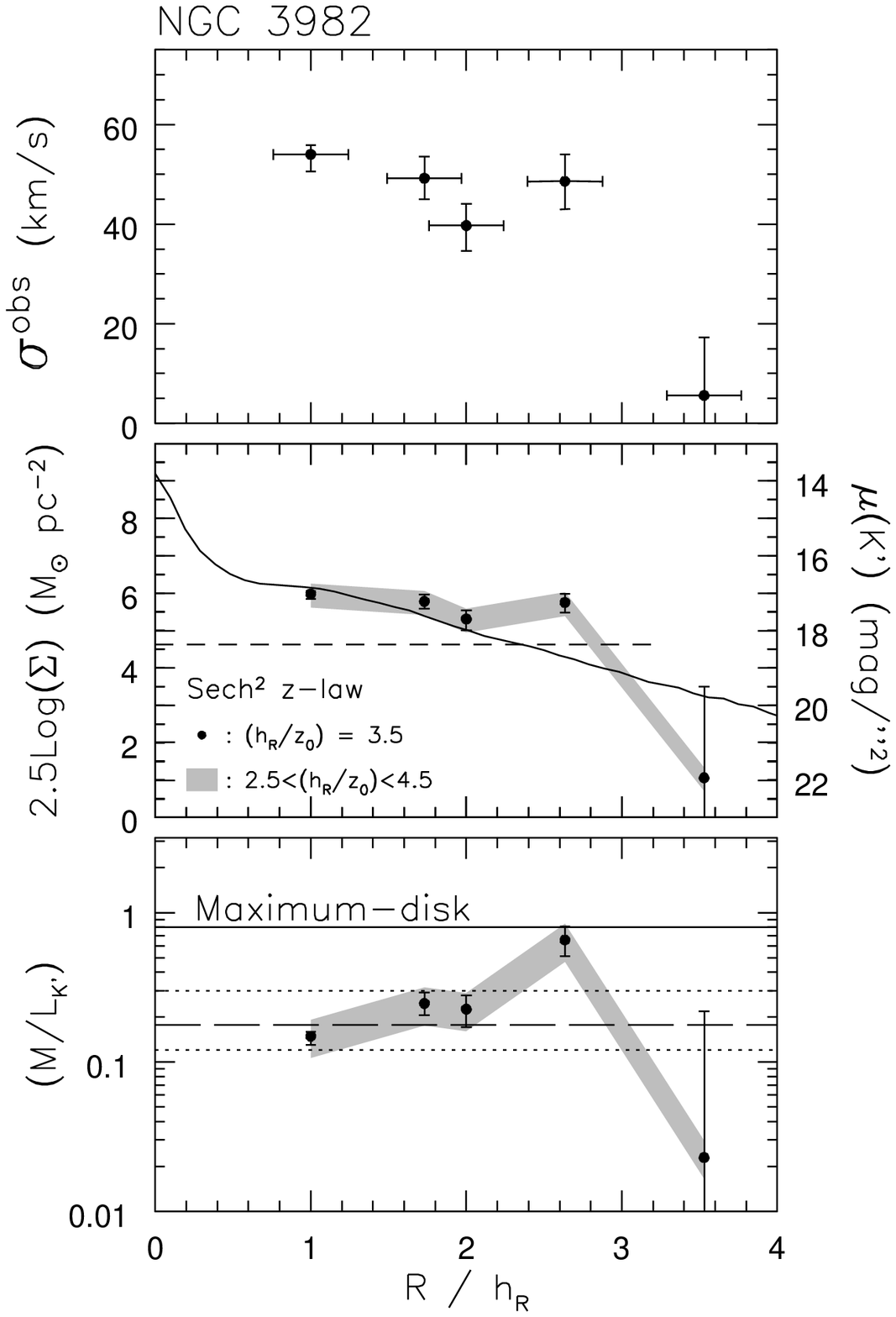}{0in}{0}{55}{55}{-75}{-160}
\vskip 1.4in
\caption{\hsize 5.2in \baselineskip 0.1625in \underline{\bf Left
panels.} \ Stellar line profiles for the deepest CaII line in N3949
(top) and N3982 (bottom). In both panels the grey curve is the
observed line for a K0.5-III template star (one of $\sim$3 dozen
template stars spanning a range in T$_{\rm eff}$ and g, observed by
drifting each star across many fibers). Major and minor axis profiles
in N3949 at R/h$_{\rm R}$ = 2 have nearly identical widths, which,
when combined with the epicycle approximation, implies
$\sigma_z\gg\sigma_R$. The radial trend in line-width in N3982 is weak
for R/h$_{\rm R}<3$. \ \underline{\bf Right panels.} \ {\it Top}:
Azimuthally averaged stellar velocity dispersions for N3982 in 5
radial bins containing 6-18 fibers (increasing with
radius). Dispersions are determined by convolving the K0.5-III stellar
spectrum with Gaussians of varying FWHM and finding the best match to
the azimuthally averaged galaxy spectra. {\it Middle}: Inferred mass
surface-density for an isothermal, vertical density profile and a
range of scale-heights. The dashed line shows the value for the Solar
Neighborhood. The solid line shows the radial K' surface brightness
profile. {\it Bottom}: Derived radial dependence of disk M/L in the K'
band. The solid line shows the maximum value allowed by the rotation
curve. The dashed line is the weighted mean over all radii while the
dotted lines map into the width of the grey, shaded curve in Figure
1. This suggests that NGC 3982 has a substantially sub-maximum disk
despite the fact that the stellar disk has a very short scale length
and an extremely high surface brightness.}
\end{figure}

Two results stand out from these initial measurements, illustrated in
Figures 2 and 3. First, for NGC 3949, the ratio of major and minor
axis velocity dispersions are unity ($\pm8$\%), while the epicycle
approximation yields $\sigma_\phi / \sigma_R = \sqrt{(1+ d {\rm ln}
V_c / d {\rm ln} R)/2} = 0.75\pm0.12$. From this we find the
surprising result that $\sigma_z$ must be much larger than either
$\sigma_R$ or $\sigma_\phi$, implying at face value that NGC 3949 has
a dynamically unusual (hot) disk (cf Gerssen et al. 2000).

Second, the velocity dispersion in the disk of NGC 3982 is $\sim$
constant with radius until the last radial datum, where $\sigma^{\rm
obs}$ falls to a significantly lower value. (No corrections were
applied for contributions from the projected radial and tangential
velocity dispersion components -- a $<$13\% effect.) We derive a {\it
total} disk mass surface density for a variety of scale heights, based
on results from recent work by Kregel et al. (2002), and assuming an
isothermal (sech$^2$) vertical density profile. This mass surface
density is at a level $\sim$3.5 higher than the surface density of the
Milky Way in the Solar Neighborhood (Kuijken \& Gilmore 1991), and is
nearly constant with radius until R/h$_{\rm R}$=3. Consequently the
mass-to-light ratio of the disk increases with radius for R/h$_{\rm
R}<3$, while the disk color gets bluer -- a trend opposite that
indicated by the models of Bell \& De Jong (2000). However, our color
range is limited, and we must also check whether red super-giants are
in abundance in this system. Finally, a weighted radial average of
(M/L$_{K^\prime}$)=0.18 is used in a rotation curve decomposition
shown in Figure 2 which implies a substantially ($\sim 25$\%)
sub-maximum, yet very high SB disk (see also Verheijen et
al. 2002). This research was supported by NSF/AST-9970780.

\begin{figure}
\plotfiddle{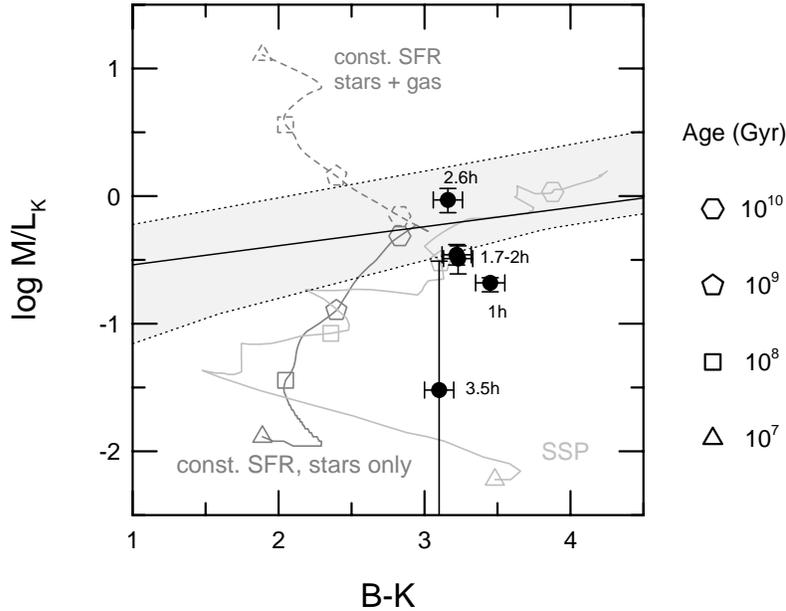}{3in}{0}{55}{55}{-140}{-75}
\caption{\hsize 5.2in \baselineskip 0.165in $K$-band mass-to-light
ratio versus B-K color for NGC 3982 at 5 radial bins (marked in units
of radial scale-length h$_{\rm R}$. The predictions from Bell \& De Jong
(2001) are shown for their full range of star-formation histories and
metallicities (grey shaded region) and favored model (solid line). The
shaded lines are single-burst (SSP) and constant star-formation models as a
function of age (see symbol key). We find NGC 3982's disk has a wide
range of M/L over a small range of B-K color.}
\vskip -0.15in
\end{figure}

\vskip -0.35in
\null

\end{document}